\documentclass[aps,prl,twocolumn,preprintnumbers,superscriptaddress,dblfloatfix,nofootinbib]{revtex4-1}
\usepackage[utf8]{inputenc}
\usepackage{lmodern}
\usepackage{amsmath,amssymb}
\usepackage{graphicx}
\usepackage{url}
\usepackage{color}
\usepackage{enumitem}
\usepackage{subfigure}
\usepackage[dvipsnames]{xcolor}
\usepackage[colorlinks=true,breaklinks=true]{hyperref}
\hypersetup{allcolors=[rgb]{0.0 0.0 0.6},linkcolor=[rgb]{0.75 0.05 0.05}}
\usepackage{bm}
\usepackage{epsfig}
\usepackage{amsmath}
\usepackage{amssymb}
\usepackage{slashed}
\usepackage{color}
\usepackage{accents}
\usepackage[dvipsnames]{xcolor}
\usepackage[colorlinks=true,breaklinks=true]{hyperref}
\hypersetup{allcolors=[rgb]{0.0 0.0 0.6},linkcolor=[rgb]{0.75 0.05 0.05}}

\newcommand{\exclude}[1]{}

\usepackage{braket}
\usepackage{url}

\usepackage{mathrsfs}

\newcommand{\tx}{\text}

\usepackage{fontawesome}

\usepackage{comment}

\newcommand{\bp}{\begin{pmatrix}}
\newcommand{\ep}{\end{pmatrix}}
\newcommand{\bb}{\begin{bmatrix}}
\newcommand{\eb}{\end{bmatrix}}

\newcommand{\df}{\text{d}}

\newcommand{\al}[1]{\begin{align}#1\end{align}}

\newcommand{\paren}[1]{\left(#1\right)}
\newcommand{\pn}[1]{\left(#1\right)}
\newcommand{\sqbr}[1]{\left[#1\right]}
\newcommand{\fn}[1]{\!\paren{#1}} 

\usepackage{fancybox}

\usepackage{amsthm}
\theoremstyle{definition}

\newcommand{\ov}{\over}

\newcommand{\ttt}{\texttt}

\newcommand{\K}{\tx{K}}

\usepackage{ulem} 

\begin{document}
\preprint{IPMU24-0019}
\preprint{RIKEN-iTHEMS-Report-24}

\title{Neutrinos and gamma rays from beta decays in an active galactic nucleus NGC 1068 jet}

\affiliation{Department of Physics and Astronomy, University of California, Los Angeles \\ Los Angeles, California, 90095-1547, USA}

\author{Koichiro Yasuda}
\affiliation{Department of Physics and Astronomy, University of California, Los Angeles \\ Los Angeles, California, 90095-1547, USA}
\author{Nobuyuki Sakai}
\affiliation{
Department of Earth and Space Science, Graduate School of Science, Osaka University, Toyonaka,
Osaka 560-0043, Japan}
\author{Yoshiyuki Inoue}
\affiliation{
Department of Earth and Space Science, Graduate School of Science, Osaka University, Toyonaka,
Osaka 560-0043, Japan}
\affiliation{Interdisciplinary Theoretical \& Mathematical Science Program (iTHEMS), RIKEN, 2-1 Hirosawa, Saitama 351-0198, Japan}
\affiliation{Kavli Institute for the Physics and Mathematics of the Universe (WPI), UTIAS \\The University of Tokyo, Kashiwa, Chiba 277-8583, Japan}
\author{Alexander Kusenko} 
\affiliation{Department of Physics and Astronomy, University of California, Los Angeles \\ Los Angeles, California, 90095-1547, USA}
\affiliation{Kavli Institute for the Physics and Mathematics of the Universe (WPI), UTIAS \\The University of Tokyo, Kashiwa, Chiba 277-8583, Japan}

\date{\today}
	
\begin{abstract}

We show that TeV neutrinos and high-energy gamma rays detected from the nearby active galaxy NGC~1068 can simultaneously be explained in a model based on the beta decays of neutrons produced in the photodisintegration of $^4$He nuclei on ultraviolet photons in the jet. The photodisintegration of nuclei occurs at energies above several PeV, which explains the 1-100~TeV energies of the observed neutrinos.  The TeV gamma-ray flux accompanying the beta decays is expected to be much lower than the neutrino flux, which agrees with the observations of NGC~1068 showing a gamma-ray deficit as compared to the expectations from proton-photon interactions. Furthermore, the synchrotron and inverse Compton gamma-ray flux associated with protons' Bethe-Heitler pair production and the photopion processes in the jet can be consistent with the observed gamma-ray flux at GeV energies for a plausible range of magnetic fields of jets. This scenario, combining beta decay and Bethe-Heitler, can be applied to other jet Seyfert galaxies such as NGC~4151. Future measurements of the neutrino flavor ratio can help confirm the beta-decay origin of the observed neutrinos. 
\end{abstract}
\maketitle
	
{\textit{Introduction--}}The IceCube discovery of TeV neutrinos from the nearby active galaxy NGC~1068 \cite{IceCube:2022der} has presented a puzzle. Neutrino production in Active Galactic Nucleus (AGN) is usually related to $p\gamma$ photopion productions, in which the gamma-ray flux becomes comparable with the neutrino flux. However, in the case of NGC~1068, the gamma ray flux observed by Fermi~\cite{Ajello:2023hkh} and MAGIC~\cite{2019ApJ...883..135A} is unexpectedly low and has an unexpected spectral shape. This discrepancy between gamma rays and neutrinos can be accommodated in a two-zone model \cite{Eichmann:2022lxh, Ajello:2023hkh}, in which the gamma rays originate from the gamma-ray thin region and the neutrinos come from the gamma-ray opaque region. For gamma rays, either star formation \cite{Eichmann:2022lxh} or AGN disk wind \cite{Peretti:2023xqk} could explain the signals depending on the model parameters. For neutrinos, hot plasma surrounding the central supermassive black hole, namely, the corona, has been discussed as a plausible production site \cite{Inoue2020, Murase2020, Eichmann:2022lxh, Neronov:2023aks, Blanco:2023dfp, Ajello:2023hkh, Mbarek:2023yeq, Fiorillo:2023dts}. However, the exact neutrino production mechanism in the corona is still unclear. Recently, it has been suggested that there is an upper bound on the cosmic ray (CR) energy budget in the Seyfert coronae, which can make it difficult to explain the NGC~1068 data~\cite{Inoue2024}.
\begin{figure*}
\centering
\begin{tabular}{cc}
    \begin{minipage}{1\columnwidth}
    \hspace{2 cm}
    \includegraphics[height = 5.5truecm]{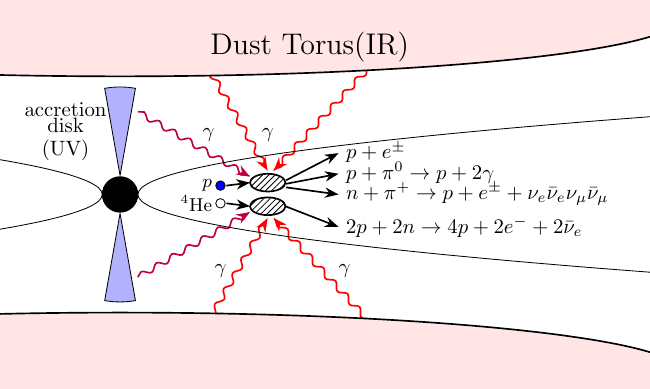}
  \end{minipage}
    \begin{minipage}{1\columnwidth}
    \hspace{1 cm}
    \includegraphics[height = 5.5truecm]{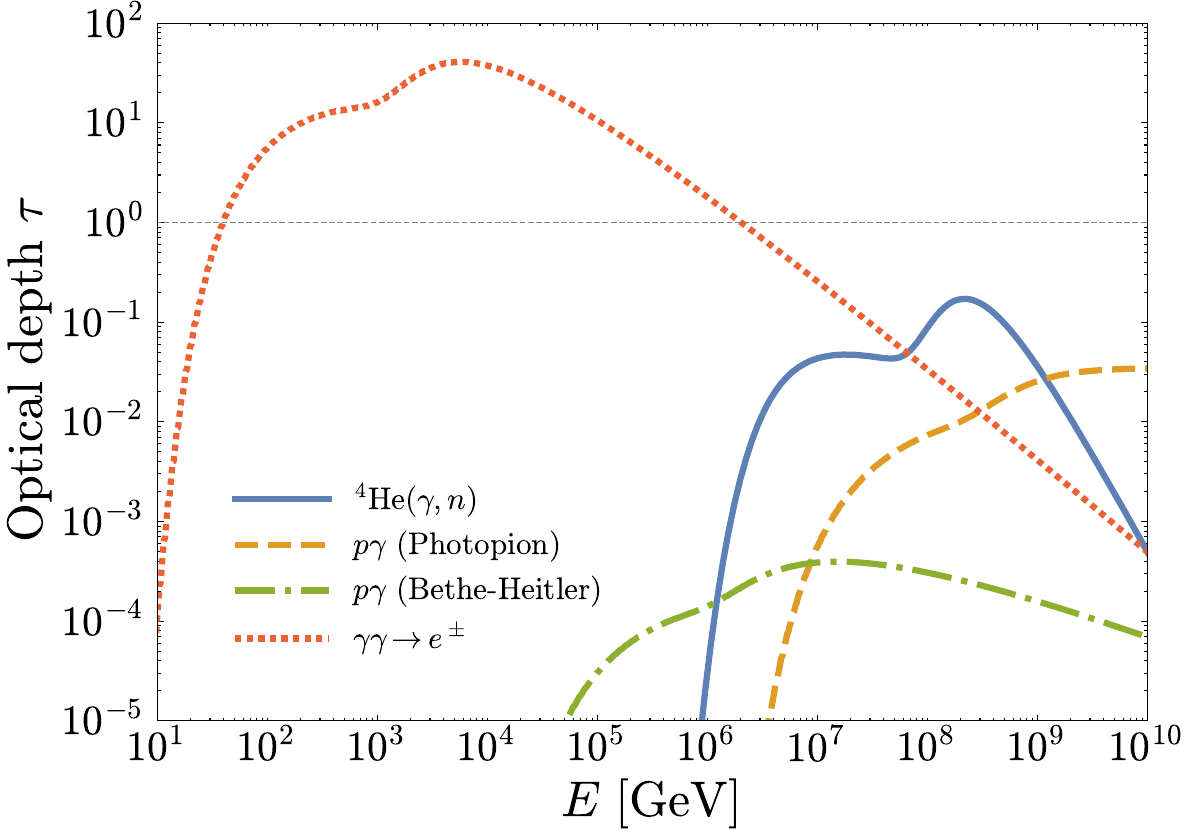}
  \end{minipage}
\end{tabular}
    \caption{($\bold{Left}$) The schematic representation of the processes, showing external radiation fields relevant for the hadronic interactions and photodisintegration. The outward open cones seen on the left and right show jets, and nuclei are accelerated within these jets. Above and below the central black hole are an accretion disk radiating in UV and a dust torus radiating in IR. Both are assumed sufficiently large compared to the scale considered, so these photon fields are approximated to be uniform and isotropic. ($\bold{Right}$) The optical depth for the photodisintegration, the photopion process, Bethe-Heitler pair production, and the $\gamma\gamma$ interaction.}
    \label{Processes}
\end{figure*}

In this {\it Letter}, we show that an alternative origin of the high-energy neutrinos and gamma rays can explain the multi-messenger observations of NGC~1068~\cite{IceCube:2022der}. Nuclei can be accelerated in the jet. A substantial fraction of the nuclei undergo photodisintegration in the ambient photon field, resulting in the release of free neutrons with high energies.  In this case, the leading source of high-energy neutrinos is not $p\gamma$ interactions but $\beta$ decays of free neutrons and daughter nuclei depending on nuclei spectra.
At the same time, protons in the jet can interact with the same photon field via $p\gamma$ and the Bethe-Heitler pair production process to create electrons and positrons.
The electrons produced by those processes interact with the ambient radiation field, such as the AGN disk photon field or the Cosmic Microwave Background (CMB) radiation, and generate high-energy gamma rays via synchrotron emission and inverse Compton (IC) scattering. We will show that the spectral shape and the flux of the resulting gamma rays and neutrinos can be in good agreement with the observations of NGC~1068.

NGC~1068 is a type-II Seyfert galaxy, an active galaxy with jet activity. 
The proton interactions in the jet cannot account for the observed GeV--TeV gamma-ray flux for several reasons. The observed diffuse X-ray luminosity at several hundred pc scales is  $L_X\sim 10^{41}$~erg~s$^{-1}$~\cite{Young2001}, and coronal X-ray luminosity at the scale of tens of Schwarzschild radius is of  $L_X\sim 10^{43}$~erg~s$^{-1}$~\cite{Koyama1989PASJ...41..731K, Bauer:2014rla, Marinucci:2015fqo}, which means there are not enough X-ray target photons in the pc-kpc scale jet for TeV neutrino production in the $p\gamma$ interactions. Additionally, even if the target photons were abundant, the fluxes of gamma rays and neutrinos expected from the $p\gamma$ interactions should be comparable, which is not supported by the data. The possible $pp$ interactions of high-energy protons on hydrogen with typical density of $n_\tx{p} = 1\tx{ cm}^{-3}$ are characterized by the time scale $\tau_\tx{pp}\sim 7\times 10^7\tx{ yrs}$, which is much longer than both the lifetime of a proton in the jet and the lifetime of the PeV jet estimated from the self-similar solution $\tau_\tx{jet}\sim 10^5\tx{ yrs}$~\cite{Michiyama2022}. Therefore, the $pp$ channel is strongly suppressed. The simplest models of proton interactions in the jet thus fail to account for the combination of the observed neutrinos and gamma rays (but see also Fang et al. \cite{Fang:2023vdg}). \\

{\textit{Photodisintegration of Nuclei--}}Let us now discuss an alternative explanation. Astrophysical jets can accelerate nuclei. The chemical composition of the ions accelerated in the jet is not known. In the observed cosmic rays, the elements appear to have abundance patterns very similar to those of the solar system~\cite{1989GeCoA..53..197A}. It is likely that the abundant stable nuclei are loaded into the jet and accelerated to PeV energies or higher (See Fig.~\ref{Processes}).

Both light and heavy nuclei present in the interstellar medium can be accelerated to energies well above TeV, but photodisintegration becomes the limiting factor to further acceleration of nuclei~\cite{Stecker:1998ib}. Photodisintegration occurs if the interacting photon has energy above the threshold energy in the rest frame of the nucleus.  The threshold varies from a few MeV for heavy nuclei to as high as 20.6~MeV for  $^4$He~\cite{1996PhDT........59R, 1999ApJ...512..521S}, which is the most abundant nucleus in the interstellar medium other than protons~\cite{1989GeCoA..53..197A}. The photodisintegration cross section of $^4$He peaks at around 25 MeV with a cross section $\sigma \approx 1.5\ {\rm mbarn}$~\cite{Quaglioni:2003ym,Shima:2005ix,Horiuchi:2012sn}.

In this work, we employ the uniform and isotropic photon field model by incorporating not only the UV black body radiation from the standard accretion disk but also thermal radiation from the dust torus, which emits predominantly in the IR range, characterized by a blackbody with temperature $10^3~\K$ (See Fig.~\ref{Processes}). Here, the standard accretion disk model typically invokes a temperature gradient, which leads to the multi-color black body radiation from the disk. These radiations from the disk exhibit a prominent feature across the optical to UV range, referred to as ``the big blue bump," emitting the most of radiation in this band. Therefore, the most dominant target photons in the jet are UV photons from the disk with energies 1-10~eV~\cite{Shakura:1972te, 2008bhad.book.....K, 1998PASJ...50..667K, Romeo2016}.
A UV photon with energy $\sim 10$~eV in the SMBH rest frame is boosted to 20.6~MeV in the $^4$He rest frame for a Lorentz factor and $^4$He ion energy 
\al{
    \gamma_N \sim 2.1\times 10^6,\quad E_{\rm He} \sim 8\ {\rm PeV}.
}
Thus $^4$He nuclei reaching the energies of $\gtrsim 8$~PeV undergo photodisintegration on UV photons producing a number of neutrons with energies $\gamma_N m_n \sim 10^6$~GeV, where $m_n$ is the neutron rest mass. This process also produces neutrinos and electrons through $\beta$ decay. The generated neutrinos have $\sim {~\rm MeV}$ energies in the frame of the neutron, and neutrino energies get boosted to the laboratory frame as
\al{
    E_\nu \sim \gamma_N\times {~\rm MeV}\sim  {\rm TeV} .    
}
We see that the TeV energy scale of the observed neutrinos is naturally explained by the photodissociation threshold of the parent nuclei accelerated in the jet. $\beta$ decays also generate electrons, which upscatter photons from the disk, torus, and CMB, producing gamma rays.  We also include the contribution of these electrons in our analysis.\\

{\textit{Model Setup--}}
We consider acceleration of protons and nuclei at the first 0.8~pc of the jet, before the jet changes its direction by an interaction with a molecular cloud \cite{Gallimore2004, Cotton2008A&A...477..517C}. We set the interaction site at $d = 0.8~$pc from the center assuming the spherical emission region with a size of $r = 0.8~$pc for simplicity. The left panel of Fig.~\ref{Processes} shows the schematic figures of the setting we consider.

The jet power of NGC~1068 was previously estimated as $10^{43}~\mathrm{erg~s^{-1}}$ based on radio measurements~\cite{Birzan2008}. However, this estimate relies on empirical relations between radio luminosity and X-ray cavity pressure established for radio galaxies~(e.g.,\cite{1999MNRAS.309.1017W, Birzan2008}), which may not directly apply to NGC~1068, a type-II Seyfert galaxy. This estimation method faces fundamental limitations: X-ray cavity pressure measurements only account for thermal components, neglecting contributions from non-thermal particles, turbulence, and magnetic fields, thus providing a lower limit. Moreover, there is also an uncertainty rin the relation between the jet power $P_\mathrm{jet}$ and the accretion power $\dot{M}c^2$. While studies of jetted AGNs using these correlations report AGN jet power $P_\mathrm{jet} \sim 0.01 \dot{M}c^2$~\cite{Inoue:2017bgt, Plsek:2022oze}, analyses of blazar spectral energy distributions (SEDs) using leptonic or leptohadronic models indicate $P_\mathrm{jet} \gtrsim \dot{M}c^2$~\cite{Ghisellini:2014pwa, Inoue:2016fwn, Rodrigues:2023vbv}. This orders-of-magnitude discrepancy underscores the substantial uncertainties in jet power determination. Given that the bolometric disk luminosity of NGC~1068 is $ = (0.4$-$4.7)\times 10^{45}~{\rm erg~s}^{-1}$~\cite{Burillo2014, 2020A&A...634A...1G}, and assuming a standard radiative efficiency of 10$\%$, the accretion power is $\dot{M}c^2 \sim 4.7\times 10^{46}~{\rm erg~s}^{-1}$. In this \textit{Letter}, we adopt the nearly maximum available power, $1.4\times10^{46}~{\rm erg~s}^{-1}$, for the jet.

The jet energy is predominantly carried by protons, with helium cosmic rays constituting a fraction determined by solar abundance ratios in terms of mass fraction \cite{1989GeCoA..53..197A}. Therefore, powers of protons and $^4$He nuclei (red) are set as $L_\tx{p}=10^{46}$~erg~s$^{-1}$ and $L_\tx{He}= 4.0\times 10^{45}$~erg~s$^{-1}$, respectively.
For NGC~1068, while jet features are apparent on various scales \cite{Gallimore2004, Michiyama2022}, observations indicate non-relativistic speeds ($<0.1c$) at scales of several parsecs from the center~\cite{Gallimore2004, Fischer2023ApJ...953...87F}, implying negligible relativistic beaming effect.

For the jet nuclei spectrum, we consider two models: a double cutoff power-law and a Maxwellian-like distribution~\cite{2011ApJ...740...64L}. The double cutoff power-law model is defined as:
\al{
    {\df N\ov \df E}\fn{E} \propto E^{-p} \exp\sqbr{-{E_\tx{low}\ov E}}\exp\sqbr{-{E\ov E_\tx{high}}},\label{doubleCutoff}
}
which creates a broad power-law spectrum  characteristic of diffusive shock acceleration \cite{1991SSRv...58..259J, 2007ApJ...667L..29S}. The Maxwellian-like distribution~\cite{2011ApJ...740...64L} is defined as
\al{
    {\df N\ov \df E}\fn{E} \propto E^2 \exp\sqbr{-\pn{E\ov E_{\rm c}}^{1 + \alpha_p}},\label{Maxwellian}.
} which yields a sharper spectral peak. This type of distribution has been suggested to explain the hard spectra observed in some blazars~\cite{2011ApJ...740...64L}. Since particle acceleration depends on magnetic rigidity, the energy cutoffs for $^4$He nuclei differ by a factor of 2 from those for protons.
 
Let us evaluate the efficiency of the relevant processes. The optical depth for each process is calculated by
\al{
	\tau_i(E) = \int d\varepsilon \frac{dn_\gamma}{d\varepsilon} \sigma_i(\varepsilon, E)r,
}
where $n_\gamma$ is the photon number density and $\sigma_i$ is the cross section of each interaction. We adopt the cross section of Photodisintegration, photopion processes, Bethe-Heitler (BH) pair production, and $\gamma\gamma$ interaction from Ref.~\cite{Dermer:2009zz, Kelner:2008ke}.
The estimated optical depth of each process is shown in the right panel of Fig.~\ref{Processes}.
We confirmed that the optical depth to photodisintegration becomes efficient at the energy of $\sim8$~PeV.
Above this energy threshold~\cite{1996PhDT........59R, 1999ApJ...512..521S}, photodisintegration of nuclei produces neutrons and protons, with subsequent neutron $\beta$ decay generating neutrinos. This photodisintegration threshold effectively also sets a lower energy cutoff for neutrino production, as discussed above.

Typically, a single $\beta$ decay produces an electron and a neutrino with similar energies in the rest frame of a neutron. In order to calculate resulting gamma-ray and neutrino fluxes, we converted those secondary distributions into the laboratory frame. The neutrino to CR energy flux ratio through the complete photodisintegration process is given by $\nu : \tx{CR} = 1 : 2\times10^3 A$~\cite{Mbarek2023}, where $A = 4$ is the mass number of $^4$He nuclei.

For blazar sub-pc scale jets, magnetic fields in the emission regions typically range from $10^{-2}$ to $10$~G~\cite{Rodrigues:2023vbv}. We adopt similar magnetic field values for our emission region. Under this assumption, the maximum attainable CR energy at distances $d\lesssim 0.8$~pc with magnetic field 3~G is estimated as $E_{\rm max}\sim q B r\approx 5\times10^{21}~{\rm eV}(q/2e)(B/3~\mathrm{G})(r/0.8~\mathrm{pc})$ following the Hillas criterion~\cite{1984ARA&A..22..425H}. This allows $^4$He nuclei to be accelerated above PeV energies, where photodisintegration becomes the dominant process as shown in Fig.~\ref{Processes}. Therefore the inner 0.8~pc of the jet is the site where the $^4$He nuclei can be accelerated to PeV energies and where the optical depth to photodisintegration becomes higher than any other hadronic processes~\cite{Quaglioni:2003ym,Shima:2005ix,Horiuchi:2012sn}.

{\textit{Secondary Gamma-ray signals--}}Let us now discuss the gamma-ray signals predicted in our model. High-energy electrons are generated not only through beta decays but also through BH pair production and photopion processes. The latter processes occur predominantly through interactions between protons and the ambient photon field. These high-energy electrons emit gamma rays through synchrotron radiation and also upscatter disk UV and torus IR photons to gamma-ray energies via inverse Compton scattering~\cite{Blumenthal1970, Aharonian1981, Khangulyan2014}.

\begin{figure*}
\centering
\begin{tabular}{cc}
    \begin{minipage}{1\columnwidth}
    \hspace{2 cm}
    \includegraphics[height = 6truecm]{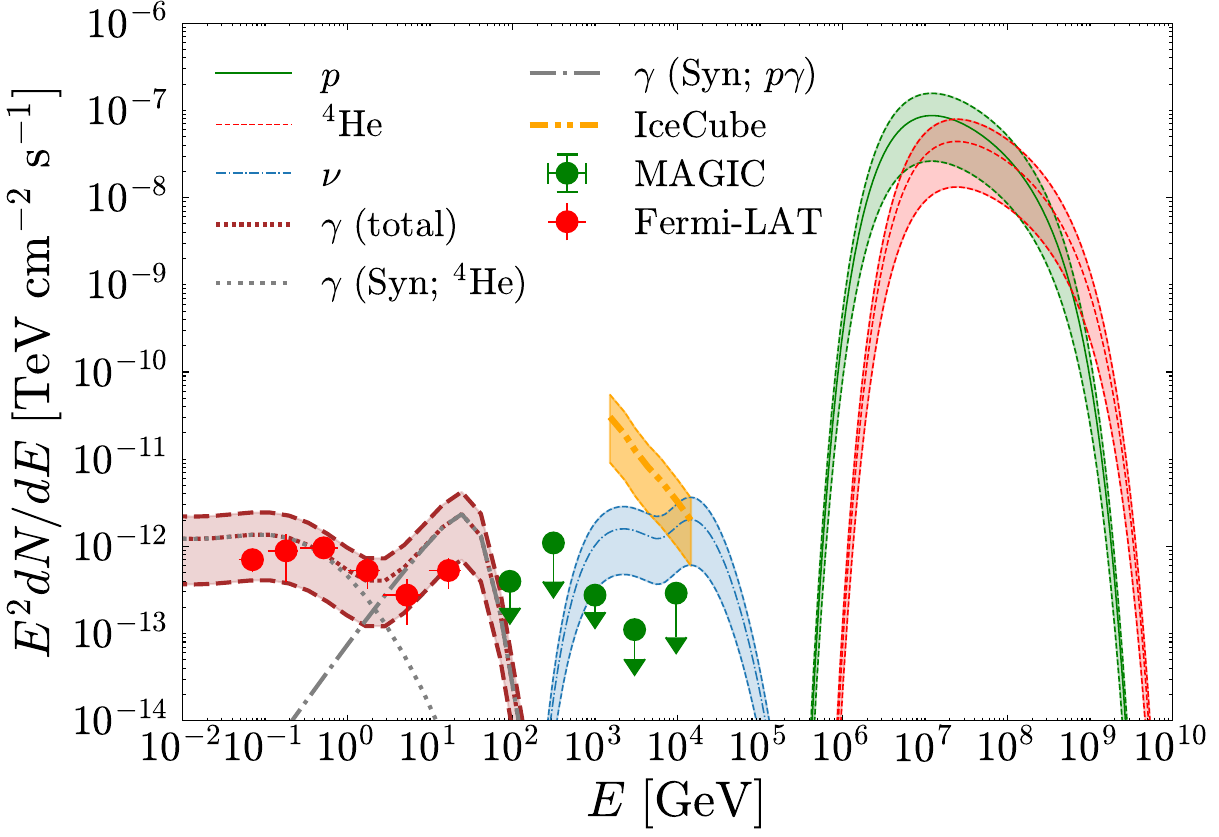}
  \end{minipage}
    \begin{minipage}{1\columnwidth}
    \hspace{1 cm}
    \includegraphics[height = 6truecm]{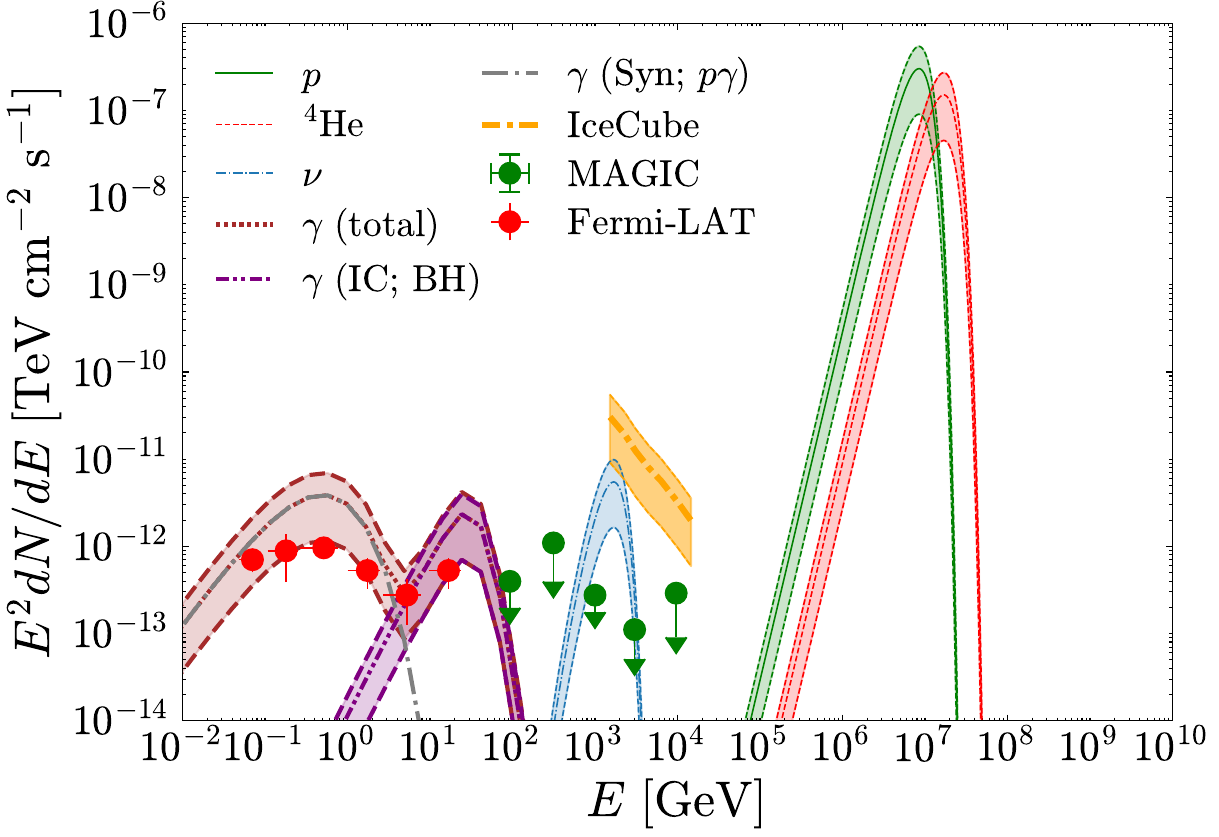}
  \end{minipage}
\end{tabular}
    \caption{Predicted neutrino and $\gamma$-ray fluxes for two models of CR spectra. ($\bold{Left}$) Double cutoff power-law scenario with the magnetic field of 3.0 G.  ($\bold{Right}$) Maxwellian-like scenario with the magnetic field of $3.0\times 10^{-2}$ G. Other parameters are discussed in the text. 
    The proton ($p$) and $^4$He fluxes at the source are displayed as indicated in the legend. The IC spectra (``$\gamma\pn{\tx{IC}}$"), synchrotron spectra (``$\gamma\pn{\tx{Syn}}$"), and the total observed gamma-ray signal, after the $\gamma\gamma$ attenuation, is also shown. The neutrino signal predicted by each model (blue band) is in general agreement with the IceCube results~\cite{IceCube:2022der} (orange band), but a detailed comparison requires a reanalysis of the IceCube data for the peaked spectrum (different from power law assumed by IceCube) and a different neutrino flavor ratio.
    The $\gamma$-ray predictions (``$\gamma\pn{\tx{total}}$") are consistent with the data from Fermi-LAT~\cite{Ajello:2023hkh} (red points) and MAGIC~\cite{2019ApJ...883..135A} (green upper bounds).}
    \label{IceCubevsBetaDecay_BHFull}
\end{figure*}

For the BH process, $e^\pm$ pairs are formed at rest in the equi-momentum reference frame, their expected Lorentz factor is given by
$\braket{\gamma_\pm} = {\gamma_{\rm p}/ \sqrt{1 + 2\varepsilon_r}}$,  
where $\varepsilon_r = {2{\gamma_\tx{p} E_\tx{disk}}/ \pn{m_e c^2}}$ is the invariant scattering energy scaled by the electron rest mass in the highly relativistic regime. For our assumed disk UV photons$\pn{\sim 10\tx{ eV}}$,  interacting with 10 PeV protons, the peak pair energy becomes~\cite{Dermer:2009zz}
\al{
    E_\pm^{\text{peak}} &= \langle \gamma_\pm \rangle m_e c^2 \simeq 17\text{ GeV},
}
which would result in $\sim100$~MeV gamma-ray photons via IC scattering of disk UV photons.

For $p\gamma$ interactions, protons having $\gtrsim10$~PeV interact with the UV photon field with a cross section of $\approx0.34$~mb with an inelasticity of $\approx0.2$ \cite{Kelner:2008ke}. This process produces PeV gamma rays and pairs, yielding a PeV gamma-ray flux of $\approx 10^{-11}~\rm{TeV~cm^{-2}~s^{-1}}$. The resulting gamma-ray emission is dominated by synchrotron radiation from these pairs depending on magnetic field strength, while their inverse Compton contribution is suppressed by the Klein-Nishina effect. Note that without additional hard X-ray photons, the electromagnetic cascade contribution to the GeV gamma-ray band remains insignificant due to synchrotron energy loss. Given our adopted jet magnetic field of $0.01\text{--}10$~G at $\sim 0.8$~pc scale, synchrotron cooling is more efficient than IC cooling.

The gamma-ray spectrum from the IC and synchrotron processes was calculated using the \ttt{naima}~\cite{Khangulyan2014,naima} Python package for relativistic particle energy distributions. For the seed photon fields, we used the AGN disk and torus photon field as described above and CMB photons. The spectral shape of the IR photon field is approximated by a single black body~\cite{2009MNRAS.394.1325R}.

{\textit{Results--}}
We present our model predictions for two different cosmic-ray spectral shapes in Fig.~\ref{IceCubevsBetaDecay_BHFull}. For both spectra, we set the luminosities to $L_\tx{p}=10^{46}$~erg~s$^{-1}$ for protons and $L_\tx{He}= 4.0\times 10^{45}$~erg~s$^{-1}$ for $^4$He nuclei.
In the double cutoff power-law model (left panel), we adopt a spectral index $p=2.6$ with energy cutoffs scaled by magnetic rigidity: ($E_{\rm low}$, $E_{\rm high}$) = (8, 200) PeV for protons and (16, 400) PeV for $^4$He nuclei. For this model, a magnetic field of 3.0 G reproduces both the observed gamma-ray and neutrino data. For the Maxwellian-like model (right panel), we use $E_{\rm c} = 7 \tx{ PeV}$ and $\alpha_p = 1.5$~\cite{2011ApJ...740...64L}, with a required magnetic field of $3.0\times 10^{-2}$ G to match the observations.

The calculated neutrino fluxes are shown in blue curves. The gamma-ray spectra show synchrotron emission (grey), inverse Compton (purple), and total emission (brown), with brown shaded regions indicating $1\sigma$ uncertainties from IceCube neutrino flux variations. Gamma rays between 10 GeV and 1 TeV are internally attenuated by target photon fields (Fig.~\ref{Processes}). Our predicted gamma-ray flux is lower than in Ref.~\cite{2024arXiv240509332D}, which assumed $B = 1~\mu\tx{G}$ and thus weaker synchrotron losses. Note that the our assumed magnetic field range ($0.01\text{--}10$ G) and region size ($\sim 0.8$ pc) are in the range of typical blazar zones~\cite{Rodrigues:2023vbv}.

The model successfully explains both TeV neutrino and gamma-ray observations from NGC~1068 through beta decay of neutrons from nuclear photodisintegration in the relativistic jet. The TeV energy scale naturally emerges from the photodisintegration threshold of nuclei (predominantly $^4$He) on disk and torus photons.

{\textit{Discussion and conclusions--}}
The observed neutrino flavor ratios can help verify our scenario. The neutrino oscillations change the flavor ratio of neutrinos propagating over cosmological distances~\cite{Fukugita:2003en, Bustamante:2015waa,Coleman:2024scd}. After traveling 14.4 Mpc~\cite{1994yCat.7145....0T, 1997Ap&SS.248....9B}, the final flavor ratio at the Earth is measurably different from that expected from the pion decays~\cite{Coleman:2024scd}, which can help confirm or rule out the beta decays as the origin of the neutrinos. 
Though IceCube can distinguish between the electron neutrinos (cascade-like events), muon neutrinos (track-like events), and tau neutrinos (cascade or double-bang events, in some energy range),
the detector has different sensitivities for the muon and electron neutrinos.
We emphasize that the published IceCube results assume a power law spectrum of high-energy neutrinos arriving with a flavor ratio of $(f_e:f_\mu:f_\tau)=(1:1:1)$, which is expected from $p\gamma$ neutrinos produced via pion decays with the flavor ratio at the source $(f_{e,S}:f_{\mu,S}:f_{\tau,S})=(1:2:0)$.
In our case, the predicted neutrino flux is a peaked speactum (not power law), and the flavor ratio at the source is  $(f_{e,S}:f_{\mu,S}:f_{\tau,S})=(1:0:0)$, resulting in the arriving flavor composition given by $f_i=\sum_{j,k}|U_{ij}|^2 |U_{kj}|^2 f_{k,S}$, leading to  $(f_e:f_\mu:f_\tau)\approx (3:1:1)$~\cite{Lipari:2007su,Anchordoqui:2014pca,Bustamante:2015waa}. Here, $U$ is the neutrino flavor mixing matrix. The IceCube data analysis based on the predicted spectral shape and flavor composition is needed to make a definitive comparison and test our model.
At present, the reported results~\cite{IceCube:2022der} appear to be consistent with both pion-decay and $\beta$-decay compositions. In the future, the flavor ratio studies could allow an additional way to verify the neutron decay origin of the neutrinos.
The neutrino flux predicted in Ref.~\cite{2024arXiv240509332D} is higher than our predictions. While it is consistent with the highest level of the range reported by IceCube, 
it results in a significantly higher jet power than what is required in our model. 

Heavier nuclei, such as, e.g., Fe, can also contribute to the signal. For heavy nuclei, the photodisintegration threshold is a few MeV (somewhat lower than that of $^4$He), and they are expected to disintegrate into lighter nuclei and individual nucleons in the jet, also starting from PeV energies. Given the high photodisintegration optical depth, the required jet power to reproduce the observed IceCube flux would be similar to that for the lighter nuclei.  The abundance of Fe in the interstellar medium is much smaller than that of helium, which is why we focused on the lighter nuclei. However, as the data quality improves, a more detailed analysis modeling the chemical composition in the jet could allow a more detailed test of the neutron decay origin of the neutrinos.

Our results can be applied to other Seyfert galaxies. If their jet structures are similar to NGC~1068, it may be possible to explain both neutrino and gamma ray spectra at the same time. Recently, it was reported that Seyfert galaxies NGC~4151 and CGCG~420-015 make a sizable contribution to the high-energy neutrino flux~\cite{IceCube:2023nai, IceCube:2023jds}. While NGC~4151 has a prominent jet~\cite{2023PASJ...75L..33I}, it is still not clear that CGCG~420-015 has an energetic jet due to its Compton-thick property\cite{2018ApJ...853..146T}. Future multi-messenger observations of  Seyfert galaxies can help understand the signals from nuclear photodisintegration and neutron decays.

We thank A.~Kochocki and  N.~Whitehorn for helpful discussions. The work of A. K. was supported by the U.S. Department of Energy (DOE) Grant No. DE-SC0009937;  by World Premier International Research Center Initiative (WPI), MEXT, Japan; and by Japan Society for the Promotion of Science (JSPS) KAKENHI Grant No. JP20H05853. YI is supported by NAOJ ALMA Scientific Research Grant Number 2021-17A; by World Premier International Research Center Initiative (WPI), MEXT; and by JSPS KAKENHI Grant Number JP18H05458, JP19K14772, and JP22K18277.

\bibliography{gammabiblio}


\end{document}